\documentclass[10pt,aps,prd,showpacs,nofootinbib,twocolumn]{revtex4-1}
\usepackage{amsfonts}
\usepackage{amssymb}
\usepackage{amsmath}
\usepackage{MnSymbol}
\usepackage{graphicx}
\usepackage{epstopdf}
\usepackage[colorlinks]{hyperref}

\date{\today}

\begin{document}

\title{Possibility of setting a new constraint to scalar-tensor theories}

\author{Raissa F.\ P.\ Mendes}
\email{rmendes@uoguelph.ca}
\affiliation{Department of Physics, University of Guelph, Guelph, Ontario, N1G 2W1, Canada}

\begin{abstract}
Scalar-tensor theories (STTs) are a widely studied alternative to general relativity (GR) in which gravity is endowed with an additional scalar degree of freedom. 
Although severely constrained by solar system and pulsar timing experiments, there remains a large set of STTs which are consistent with all present day observations.
In this paper, we investigate the possibility of probing a yet unconstrained region of the parameter space of STTs based on the fact that stability properties of highly compact neutron stars in these theories may radically differ from those in GR. 
\end{abstract}


\maketitle

\section{Introduction}
\label{sec:intro}

From the weak post-Newtonian regime to the strong field environment of neutron stars, general relativity (GR) has by now passed a variety of tests \cite{Will2006}. In this effort of probing GR's domain of validity, it is essential to understand how its predictions differ from those of larger classes of theories of gravity, so that observational data can be fully used to restrict the parameter space of possible models. In particular, a well-motivated and largely studied extension of GR consists of scalar-tensor theories (STTs) \cite{Damour1992,Fujii2003}, in which gravity is mediated by one or more scalar fields in addition to the usual spin-2 field. For one additional scalar field, the general action reads \cite{Sotiriou2008}
\begin{align}
	S &= \frac{1}{16\pi} \int{d^4 x \sqrt{-g} \left[ A(\phi) R 
	- 2 B(\phi) \nabla_\mu \phi \nabla^\mu \phi 
	- V(\phi)\right]} \nonumber \\
	&+ S_m\left[\Psi_m ; e^{2 a(\phi)} g_{\mu\nu} \right],
\label{general_action}
\end{align}
where the integral term is the gravitational action, written in terms of the metric $g_{\mu\nu}$ and the scalar field $\phi$, and the second term is the matter action, which is a functional of the matter fields, collectively denoted by $\Psi_m$, and of the metric $\bar{g}_{\mu\nu} \equiv e^{2 a (\phi)} g_{\mu\nu}$. 
The formal invariance of Eq.~(\ref{general_action}) under the field redefinitions $\phi \to f(\phi)$ and $g_{\mu\nu} \to e^{2 b(\phi)} g_{\mu\nu}$ allows one to fix two of the arbitrary functions $A,B,V$ and $a$ without restricting the underlying physical theory; by choosing the remaining two, one specifies a particular theory within this class. 

Decades of active research into the phenomenology of STTs (from the early works on Jordan-Brans-Dicke theory \cite{Jordan1959,Brans1961} to the influential papers of Damour and Esposito-Far\`ese \cite{Damour1993,Damour1996} and many modern developments), together with precise solar system and pulsar timing experiments, were able to put important bounds on the parameter space of STTs. In particular, in the ``Einstein frame'' ($A = B = 1$), theories with $V = 0$ can be described by a set of parameters $\{\alpha_0,\beta_0,\dotsc\}$ defined through the expansion $a(\phi) = \alpha_0 (\phi - \phi_0) + \frac{1}{2} \beta_0 (\phi - \phi_0)^2 + \dotsb$ around a background value $\phi_0$ of the scalar field. The current observational bounds on the subset $\{\alpha_0,\beta_0\}$ are summarized, e.g., in Fig.~7 of Ref.~\cite{Freire2012}. Since the post-Newtonian parameters which characterize deviations of STTs from GR are all proportional to powers of $\alpha_0$ \cite{Damour1996a}, solar system experiments, such as the measurement of Shapiro time-delay variations \cite{Bertotti2003}, mainly constrain this parameter, and, for $|\alpha_0|$ sufficiently small, these tests are satisfied independently of the remaining $\{\beta_0, \dotsc\}$. Tests in the nonperturbative strong field regime are then needed in order to further constrain the parameter space of STTs. In particular, it was shown in Ref.~\cite{Damour1993} 
that relativistic stars in STTs can differ significantly from those in GR by the presence of a nontrivial scalar field profile, which could leave observable imprints, most notably, in pulsar timing observables \cite{Damour1996}. 
This ``spontaneous scalarization'' phenomenon is known to take place for arbitrarily small values of $|\alpha_0|$, and for\footnote{The precise value depends on the equation of state (see, e.g., Ref.~\cite{Shibata2014}), and can be also modified by rotation \cite{Doneva2013,Mendes2014b}.} $\beta_0 \lesssim -4.35$ \cite{Harada1998}. 
Due to the existence of this strong field effect in STTs, the analysis of pulsar timing data has been able to exclude values of $\beta_0 \lesssim -4.5$ \cite{Freire2012}, providing the best current constraints to STTs. 
Scalarization can also play an important role in the dynamics of tight neutron star binaries \cite{Barausse2013,Shibata2014,Palenzuela2014}, with observable signatures in the emitted gravitational waves.

In this paper, we explore the possibility of using measurements of neutron stars in order to probe STTs with $\beta_0 >0$ (and arbitrarily small $|\alpha_0|$), and thereby further restrict the parameter space of these theories. This possibility relies on the fact that, for some realistic equations of state (EOS), sufficiently compact neutron stars which are stable according to GR are no longer stable in some STTs with $\beta_0>0$. 
Indeed, it has long been known \cite{Harada1997} that if the trace $T$ of the energy-momentum tensor of the stellar fluid acquires a positive value somewhere inside the star, this would lead to an instability if the underlying gravitational theory were a STT with a sufficiently large value of $\beta_0>0$. 
However, as far as we know, the implications of this instability have not yet been analyzed in detail, maybe because this condition on $T$ was thought to be ``unnatural'' for ordinary matter. 
Therefore, in this paper we use the fact that, above nuclear density, matter may satisfy this condition \cite{Haensel2007} to analyze how astrophysical observations of neutron stars could in principle be used to impose new constraints on STTs.

The paper is organized as follows. In Sec.~\ref{sec:preliminars} we briefly review the linear stability analysis of spherically symmetric stars in STTs, focusing on those with $\beta_0>0$. 
In Sec.~\ref{sec:instability} we investigate how the conditions that lead to an instability of stellar configurations in STTs with $\beta_0>0$ depend on the EOS of the neutron star, and explore the parameter space of the instability for a particular set of realistic EOS. In Sec.~\ref{sec:scalarization}, we discuss the relation of the instability to the phenomenon of spontaneous scalarization, and we draw our main conclusions in Sec.~\ref{sec:conclusions}. We adopt units in which $c=G=1$ in the main equations, but restore cgs units when appropriate.

\section{Stellar Stability in Scalar-Tensor Theories}
\label{sec:preliminars}

We begin by briefly reviewing the linear stability analysis of spherically symmetric stars in STTs \cite{Harada1997}. Variation of Eq.~(\ref{general_action}) in the Einstein frame ($A = B = 1$) yields the field equations
\begin{align} 
	G_{\mu\nu} - 2 \nabla_\mu \phi \nabla_\nu \phi &+ 
	g_{\mu\nu} \nabla_\rho \phi 
	\nabla^\rho \phi + \frac{1}{2} V(\phi) g_{\mu\nu}
	= 8\pi T_{\mu\nu}, \label{g_eq} \\
	\nabla_\mu \nabla^\mu \phi &- \frac{1}{4} 
	\frac{d V(\phi)}{d\phi} = -4 \pi \frac{da(\phi)}{d\phi} T,
	\label{phi_eq}
\end{align}
which, together with the equations of motion for the matter fields,
\begin{equation} \label{eqs_motion}
	\nabla_\mu T^\mu_{\;\;\nu} = \frac{da(\phi)}{d\phi} T \nabla_\nu \phi, 
\end{equation}
determine the system dynamics. Here, all geometric quantities refer to the Einstein-frame metric $g_{\mu\nu}$, and the energy-momentum tensor of the matter fields in this frame is defined as $T^{\mu\nu} \equiv 2 (-g)^{-1/2} \delta S_m [\Psi_m ; e^{2a(\phi)} g_{\rho\sigma} ]/\delta g_{\mu\nu}$; also, $T \equiv T^{\mu\nu} g_{\mu\nu}$. 

In the following, we restrict our attention to the special case $V(\phi) = 0$, since we are mainly interested in the effects of the nonminimal coupling induced by the function $a(\phi)$, which again can be expanded as
\begin{equation} \label{a}
	a(\phi) = \alpha_0(\phi - \phi_0) + \frac{1}{2} \beta_0 (\phi - \phi_0)^2 + \dotsb.
\end{equation}
Here, $\phi_0$ is a constant background value of the scalar field that is determined by cosmological boundary conditions. In fact, for $\beta_0 > 0$, it can be shown \cite{Damour1993b,Damour1993a} that cosmological evolution naturally drives the scalar field towards a minimum of $a(\phi)$, which is achieved at $\phi = \phi_0$ if $\alpha_0 = 0$. Since $\beta_0 > 0$ is the case in which we are most interested, and since observational constraints are all consistent with $\alpha_0 \approx 0$ \cite{Freire2012}, we assume the condition $\alpha_0 = 0$ to hold and study the linear stability of equilibrium configurations for which $\phi = \phi_0$. It can be readily verified that, for a constant value of the scalar field, Eq.~(\ref{phi_eq}) is trivially satisfied, while Eqs.~(\ref{g_eq}) and (\ref{eqs_motion}) reduce to the general relativistic form
\begin{align}
	&G_{\mu\nu} = 8\pi T_{\mu\nu} \label{E_eq}\\
	&\nabla_\mu T^{\mu\nu} = 0. \label{T_eq}
\end{align}
Let us therefore investigate the linear stability of the configuration consisting of a constant scalar field and an equilibrium solution of Eqs.~(\ref{E_eq})-(\ref{T_eq}) with respect to a small perturbation of the scalar field, $\phi = \phi_0 + \delta \phi$, which can have either classical or quantum \cite{Lima2010,Mendes2014} origins. 
Since Eqs.~(\ref{g_eq}) and (\ref{eqs_motion}) depend quadratically on $\delta \phi$ [for $\alpha_0 = V(\phi) = 0$], we see that the changes in the spacetime geometry and in the motion of matter fields engendered by $\delta \phi$ vanish to first order, so that the scalar field perturbation effectively evolves on a fixed background:
\begin{equation} \label{perturbation_eq}
	\nabla_\mu \nabla^\mu \delta \phi = - 4 \pi \beta_0 T \delta \phi,
\end{equation}
where covariant derivatives refer to the background metric, which is a solution of Eq.~(\ref{E_eq}). In particular, let us consider an equilibrium solution of Eq.~(\ref{E_eq}) with spherical symmetry,
\begin{equation} \label{metric}
	ds^2 = - e^{2\Xi (r)} dt^2 + e^{2\Lambda (r)} dr^2 + r^2
	(d\theta^2 + \sin^2 \theta d\varphi^2),
\end{equation}
which is sourced by a self-gravitating perfect fluid, with energy-momentum tensor
\begin{equation} \label{PF}
	T^{\mu\nu} = \epsilon u^\mu u^\nu + p (g^{\mu\nu} + u^\mu u^\nu),
\end{equation}
where $u^\mu$ is the fluid four-velocity and $\epsilon$ and $p$ are the energy density and pressure, respectively, as measured in the fluid rest frame. (Note that, in this linear regime, the metric and matter energy-momentum tensor are the same both in the ``Einstein'' and ``Jordan'' \cite{Sotiriou2008} conformal frames.) 
Provided an equation of state $p = p(\epsilon)$ for cold matter and the value of the pressure at the center of the star, Eqs.~(\ref{E_eq})-(\ref{T_eq}) for a perfect fluid (\ref{PF}) in spherical symmetry (\ref{metric}) can be evolved as usual, determining the spacetime and matter distribution. Then, a solution to Eq.~(\ref{perturbation_eq}) in this fixed background can be written as a combination of modes of the form
\begin{equation} \label{delta_phi}
	\delta \phi_{\omega lm} (t,r,\theta,\varphi) = e^{-i\omega t} \frac{\psi_{\omega l}(r)}{r} 
	Y_{lm}(\theta,\varphi),
\end{equation}
where $\psi_{\omega l} (r)$ obeys
\begin{equation} \label{diff_eq}
	\left( -\frac{d^2}{dx^2} + V_\textrm{eff}^{(l)}(r)\right) \psi_{\omega l}[r(x)] = \omega^2 \psi_{\omega l}[r(x)],
\end{equation}
with $x(r) \equiv \int_0^r e^{\Lambda(r')-\Xi(r')} dr'$ and the effective potential given by
\begin{equation} \label{Veff}
	V_\textrm{eff}^{(l)}(r) =  \frac{e^{2(\Xi-\Lambda)}}{r} \left( \frac{d\Xi}{dr} - \frac{d\Lambda}{dr}\right) + \frac{l(l+1)}{r^2}e^{2 \Xi} - 4\pi \beta_0 T e^{2 \Xi}.
\end{equation}

If the last term in Eq.~(\ref{Veff}) is non-negative, the operator $-d^2/dx^2 + V_\textrm{eff}^{(l)}(x)$ in Eq.~(\ref{diff_eq}) has a purely positive spectrum \cite{Mendes2014}; however, if this last term is sufficiently negative, the same operator may have additional negative eigenvalues $\omega^2 = -\Omega^2 <0$, the presence of which implies that the equilibrium configuration is mode unstable under scalar field perturbations. Note that, if Eq.~(\ref{metric}) describes the spacetime of a relativistic star with radius $R$, then the time scale of the instability is roughly $\tau = \Omega^{-1} \sim R$, which is of the order of milliseconds for a typical neutron star. (We refer to Refs.~\cite{Harada1997,Mendes2014} for more details on this linear stability analysis.)

For $\beta_0 < 0$, this instability has been recognized as the linear manifestation of the spontaneous scalarization phenomenon \cite{Harada1997,Harada1998,Pani2011}, so that the unstable configuration with $\phi = \phi_0$ would be driven towards a scalarized equilibrium state with a nontrivial profile of the scalar field (see, e.g., Ref.~\cite{Novak1998} for a numerical study of this transition). However, as we discuss in more detail in Sec.~\ref{sec:scalarization} (see also Ref.~\cite{Pani2011}), a similar stabilization mechanism may not be available when $\beta_0 >0$. In this case, equilibrium configurations which would be stable according to GR may be prevented from forming in some STTs, and by observing such stars it would be possible to further constrain these alternative theories of gravity. 

Before proceeding to a more detailed analysis, some general comments are in order. In STTs characterized by some $\beta_0>0$, a necessary condition for the instability of stellar configurations under perturbations of a constant scalar field background is that the trace $T$ of the energy-momentum tensor of the stellar fluid be positive in some region inside the star, so that the effective potential (\ref{Veff}) can be negative. Moreover, if $T>0$ somewhere inside the star, the instability is always triggered for sufficiently large values of $\beta_0$.
In the case of a perfect fluid, the condition on the trace reads
\begin{equation} \label{T}
	T(r) = 3p(r) - \epsilon(r) > 0, \quad \textrm{for some } r \in [0,R].
\end{equation}
For free noninteracting particles the relation $p \leq \epsilon/3$ holds, with the equality corresponding to the limit of ultrarelativistic motion. However, although the condition $T \leq 0$ is commonly satisfied, it is not of a fundamental nature, and can be violated by strongly interacting systems without implying the breakdown of any physical principle, such as causality (this was exemplified long ago by Zel'dovich \cite{Zeldovich1962}). In particular, condition (\ref{T}) could in principle be satisfied inside neutron stars. As a preliminary example, we can take the limiting case of a maximally stiff equation of state ($dp/d\epsilon = 1$), which is constrained only by causality (see, e.g., Ref.~\cite{Koranda1997}):
\begin{equation}\label{stiff}
	p = \epsilon - \epsilon_0 \textrm{ for } \epsilon > \epsilon_0, 
 \qquad p = 0 \textrm{ for } \epsilon < \epsilon_0.
\end{equation}
The general relativistic equations of hydrostatic equilibrium can be solved after a central value $\epsilon_c$ for the energy density has been specified. However, GR sets a maximum value for the central energy density above which stars are unstable against gravitational collapse; for Eq.~(\ref{stiff}), this is $\epsilon_c = 3.03 \epsilon_0$, which corresponds to a compactness $M/R = 0.354$. On the other hand, from Eqs.~(\ref{T})-(\ref{stiff}), we have $T = \epsilon_0 (2 \epsilon/\epsilon_0 -3)$, which is positive in some region inside the star if its central energy density exceeds $3\epsilon_0/2$ or, equivalently, if its mass-to-radius ratio satisfies $M/R > 0.286$. Therefore, condition (\ref{T}) can indeed be satisfied by configurations which are both causal and stable\footnote{Throughout the paper, ``stability'' in GR refers to linear thermodynamic stability, identified by means of a turning point criterion in a sequence of equilibrium configurations \cite{Sorkin1981}.}
according to GR.
However, whether this condition is realized in Nature depends crucially on the equation of state for neutron stars, which is still subject to much uncertainty. In the following section, we investigate how generic this condition is and show that it can indeed be satisfied by neutron stars obeying realistic EOS. The implications for STTs are then analyzed for a set of these EOS.

\section{EOS Dependence and Parameter Space of the Instability}
\label{sec:instability}

\subsection{Parametrizing the EOS}

In the core of neutron stars, densities up to several times the nuclear saturation density ($\sim 2.7 \times 10^{14}$ g$/$cm$^3$) can be reached; however, our current understanding of the microscopic behavior of matter under such extreme conditions is still rather poor. Moreover, although the EOS of neutron stars can in principle be reconstructed from precise measurements of their masses and radii \cite{Lindblom1992}, in practice it can be extremely hard to infer these quantities (specially the radius) from electromagnetic observations (see, e.g., Ref.~\cite{Steiner2010}). Since there remain many uncertainties on the EOS of neutron stars, it is often convenient to provide a phenomenological parametrization for it, in order to study the implications of current constraints in a generic way. In particular, in Ref.~\cite{Read2009} the EOS was modeled as a four-parameter piecewise polytrope, and it was shown that the best fit to 34 theoretical candidate EOS using this parametrization could reproduce the properties predicted by those models with good accuracy. Since we will adopt this parametrization in the following, let us briefly review how it is constructed. For each rest-mass density interval $\rho_{i-1} \leq \rho \leq \rho_i$ of the piecewise polytrope, the pressure satisfies
\begin{equation}
	p(\rho) = K_i \rho^{\Gamma_i},
\end{equation}
where $\Gamma_i$ is the adiabatic index and $K_i$ is chosen in order to ensure continuity at each boundary. The energy density is then determined from integrating the first law of thermodynamics
\begin{equation}
	d \frac{\epsilon}{\rho} = -p d \frac{1}{\rho}.
\end{equation}

\begin{figure*}[t]
\centering
\includegraphics[width=17cm]{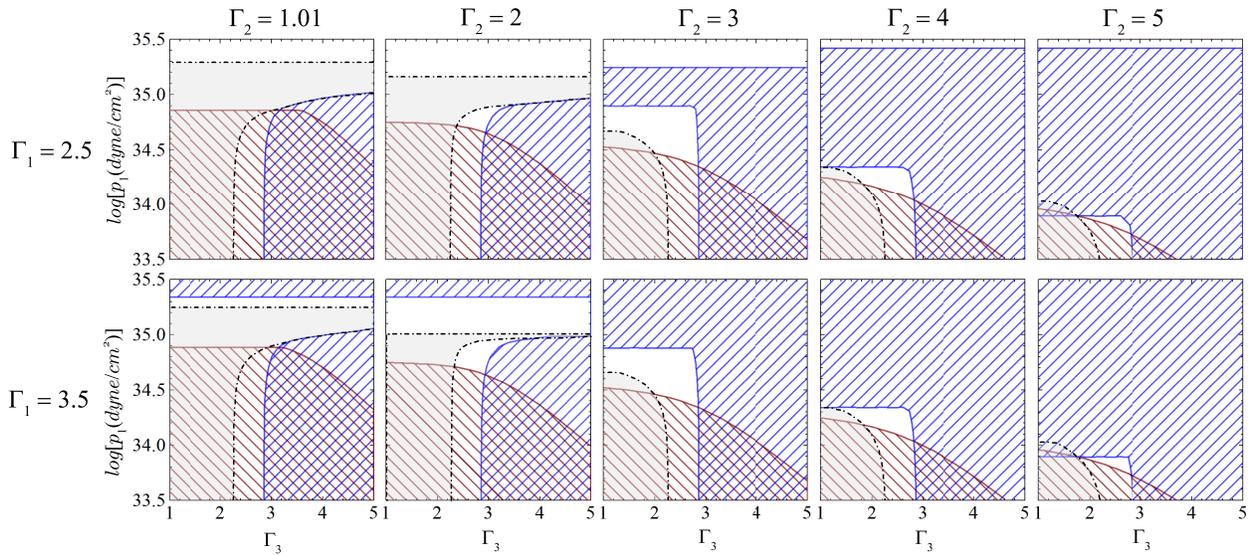}
\caption{For some slices in the space of piecewise polytropic EOS ($\Gamma_1 \in \{2.5, 3.5\}$, and $\Gamma_2 \in \{1.01, 2, 3, 4, 5\}$), we show the regions of EOS that fail to provide (i) a maximum neutron star mass of $2 M_\odot$ (filled with dark red descending diagonal lines), and (ii) a subluminal speed of sound inside stable stars (blue ascending diagonal lines). We also represent, in light gray, the regions of EOS for which $T\leq0$ everywhere inside stable stars, and outside of which positive values of $T$ can be reached. The solid white regions thus correspond to EOS which satisfy both the constraint imposed by causality and by a sufficiently high maximum neutron star mass, and at the same time allow for the condition $T>0$ to be satisfied inside some stable stars.}
\label{fig:EOSspace}
\end{figure*}

Specifically, the authors fix the EOS at low densities (adopting the version of Ref.~\cite{Douchin2001}), and match it to a polytrope with adiabatic index $\Gamma_1$. At a fixed density $\rho_1 = 10^{14.7}$ g$/$cm$^3$ and pressure $p_1=p(\rho_1)$, the EOS is then matched to a second polytropic phase, with adiabatic index $\Gamma_2$, and, at a density $\rho_2 = 10^{15}$ g$/$cm$^3$, a third phase with adiabatic index $\Gamma_3$ ensues. The set of parameters $\{\log(p_1),\Gamma_1,\Gamma_2,\Gamma_3\}$ that best fits several theoretical EOS is given in Table III of Ref.~\cite{Read2009}, to which we refer the reader for more details on this parametrization.

\subsection{Trace condition}
\label{sec:trace}

Let us consider the space of EOS defined by $33.5 < \log(p_1/($dyne cm$^{-2})) \leq 35.5$, $1.4 < \Gamma_1 \leq 5$, $1 < \Gamma_2 \leq 5$, and $1 < \Gamma_3 \leq 5$, which encompasses all 34 EOS studied in Ref.~\cite{Read2009}, and investigate the genericness of the condition that the EOS allows for configurations in which Eq.~(\ref{T}) is satisfied. 

In Fig.~\ref{fig:EOSspace}, we show some slices of this space of piecewise polytropic EOS. For each EOS, we compute the maximum central density $\rho_\textrm{c,max}$ of stars that are stable 
in GR, by means of a turning point criterion \cite{Sorkin1981} (but accommodating cases in which there is an ``instability island'' due to a softening of the EOS in the intermediate polytropic region \cite{Read2009}).
In Fig.~\ref{fig:EOSspace} the regions of EOS for which $T(\rho)\leq 0$ for $\rho \leq \rho_\textrm{c,max}$ are shown in light gray, and the complementary regions are those for which $T(\rho)>0$ for some $\rho \leq \rho_\textrm{c,max}$. 
We also represent the regions which are excluded by the constraints that (i) the EOS must allow for a maximum mass greater than $2 M_\odot$, which is approximately the highest measured neutron star mass \cite{Demorest2010,Antoniadis2013}, and (ii) that the speed of sound $v_s = \sqrt{dp/d\epsilon}$ must not surpass the speed of light, as demanded by causality, for stars in the stable sequence. Although these conditions severely constrain the space of EOS, we see that there is a considerable region in this space allowing for configurations inside of which $T$ assumes positive values.

\begin{figure}[ht]
\includegraphics[width=9.5cm]{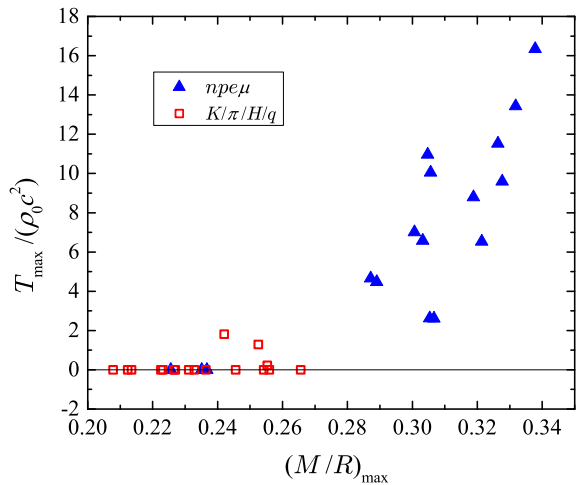}
\caption{For each polytropic approximation to the 34 EOS studied in Ref.~\cite{Read2009}, we represent $T_\textrm{max} \equiv \max_{0\leq\rho\leq\rho_\textrm{c,max}} T(\rho)$ (in units of $\rho_0 c^2$, where $\rho_0 \equiv 2.7 \times 10^{14}$ g/cm$^3$), where $\rho_\textrm{c,max}$ is the highest central density of a stable star in each model, as a function of the maximum compactness allowed by each EOS. Blue triangles correspond to EOS modeling plain $npe\mu$ nuclear matter, while red squares correspond to models including hyperons, pions, kaons or quarks. Positive values of $T$ are generally achieved inside stars with large compactness.}
\label{fig:Tmax}
\end{figure}

Next, let us consider the piecewise polytopic approximation to the 34 EOS studied in Ref.~\cite{Read2009} (cf.~Table III of that paper). These are EOS generated by several different methods and which include the effects of different particle species. In Fig.~\ref{fig:Tmax}, we show, for each of these approximated EOS, a point $(M/R,T)_\textrm{max}$, corresponding to the maximum value of $T(\rho)$ for $0\leq\rho\leq\rho_\textrm{c,max}$, 
 as a function of the compactness $M/R$ of a star with central density $\rho_\textrm{c,max}$. We distinguish equations of state which include only nucleons and leptons (the so-called $npe\mu$ models) from those including hyperons, pion and kaon condensates, and quarks ($K/\pi/H/q$ models), by different symbols.
Out of these 34 EOS, we find that 17 of them allow for $T(\rho)>0$ for some $\rho\leq \rho_\textrm{c,max}$. The points corresponding to the remaining EOS gather at $T_\textrm{max} = T(\rho)|_{\rho=0} = 0$, which is always (trivially) achieved at the radius of the star.
From Fig.~\ref{fig:Tmax}, we see that stiff equations of state allowing for more compact stars are those for which a positive value of $T$ is generally found. These are characteristic of most models describing plain $npe\mu$ matter.
On the other hand, since the presence of hyperons, pions, kaons and quarks in general leads to softer EOS, and less compact stars, we see that it tends to disfavor a positive value of $T$. 

Finally, note that, out of the 34 EOS studied in Ref.~\cite{Read2009}, only seven of them satisfy the constraints imposed by causality and the observation of a neutron star with mass around $2 M_\odot$, namely SLy, ENG, MPA1, MS1, MS1b, H4 and ALF2. The first five, which describe plain $npe\mu$ nuclear matter, allow for stable stars in which the trace of the fluid's energy-momentum tensor can take positive values, while the last two, which include either hyperons (H4) or quark matter (ALF2), are softer and do not allow for such configurations. Thus, in the following, we focus on the SLy, ENG, MPA1, MS1 and MS1b models in order to investigate in more detail the parameter space of the instability described in Sec.~\ref{sec:preliminars}.

\subsection{Parameter space of the instability}
\label{sec:paramspace}

In Sec.~\ref{sec:preliminars} we argued that, although stellar equilibrium configurations in STTs can be very close to GR's, their stability properties can differ. In fact, in Sec.~\ref{sec:trace} we saw that some realistic EOS allow for stars which are stable according to GR, but which lead to an instability in STTs with sufficiently large $\beta_0 >0$.
In this subsection, we focus on neutron stars described by piecewise polytropic approximations to the SLy, ENG, MPA1, MS1 and MS1b models (see Table I). Then, in these backgrounds, we look for unstable solutions of Eq.~(\ref{diff_eq}), i.e., solutions with $\omega^2 = -\Omega^2 < 0$ and subject to the boundary conditions
\begin{equation}
	\psi_{\Omega l}[r(x)]_{x=0}=0, \quad \psi_{\Omega l}[r(x)]_{x\to \infty} \sim e^{-\Omega x} \quad (\Omega > 0),
\end{equation}
which follow from Eq.~(\ref{diff_eq}) by requiring that the mode (\ref{delta_phi}) is everywhere regular and well behaved. This search is done by means of a shooting method, in which, for some $\Omega>0$, partial solutions of Eq.~(\ref{diff_eq}) obeying the correct boundary conditions inside and outside the star are matched into a continuously differentiable solution on the stellar surface by adjusting the remaining $\beta_0$ parameter.

\begin{table}[bth]
\begin{tabular}{ | l | c | c | c | c | c | c | c | }
	\hline                       
	EOS & $\log(p_1)$ & $\Gamma_1$ & $\Gamma_2$ & $\Gamma_3$ & $\left(\frac{M}{R}\right)_\textrm{max}$ & $\left(\frac{M}{R}\right)_\textrm{min}$ & $\beta_0^\textrm{cr}$ \\
	\hline
	SLy & 34.384 & 3.005 & 2.988 & 2.851 & 0.303 & 0.262 & 20.6 \\
	ENG & 34.437 & 3.514 & 3.130 & 3.168 & 0.319 & 0.264 & 11.1 \\
	MPA1 & 34.495 & 3.446 & 3.572 & 2.887 & 0.321 & 0.265 & 12.3 \\
	MS1 & 34.858 & 3.224 & 3.033 & 1.325 & 0.305 & 0.265 & 24.8 \\
	MS1b & 34.855 & 3.456 & 3.011 & 1.425 & 0.307 & 0.267& 25.4 \\
	\hline  
\end{tabular}
\caption{We show the set of polytropic parameters $\{\log(p_1),\Gamma_1,\Gamma_2,\Gamma_3\}$ ($p_1$ in dyne$/$cm$^2$) that give the best fit to five realistic EOS, as computed in Ref.~\cite{Read2009}. For each piecewise polytropic approximated EOS, we present (i) the maximum compactness $(M/R)_\textrm{max}$ of stable stars in GR, (ii) the ``minimum compactness'' $(M/R)_\textrm{min}$ below which stars in each model have $T \leq 0$ everywhere, and (iii) the critical value of $\beta_0$ such that, for STTs with $\beta_0 > \beta_0^\textrm{cr}$, there would be stars which are stable according to GR but unstable in that STT.}
\label{table:EOS}
\end{table}

\begin{figure}[tbh]
\includegraphics[width=8.5cm]{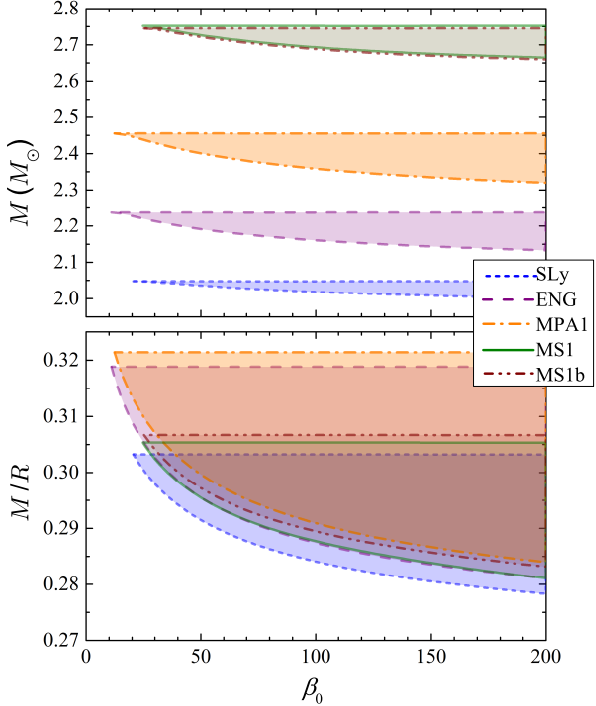}
\caption{Shaded regions are plotted in the $(\beta_0,M)$ and $(\beta_0,M/R)$ planes corresponding to configurations in which unstable solutions for the scalar field perturbation are found. Different regions (and colors) are associated with distinct EOS, and are cut at the maximum mass allowed by GR for each model (or corresponding compactness).}
\label{fig:instability}
\end{figure}

In Fig.~\ref{fig:instability}, we show the regions in the parameter space $(\beta_0,M)$ or $(\beta_0,M/R)$ in which unstable solutions for the scalar field perturbation are found. Each shaded region corresponds to a choice of EOS, and only stellar models that are stable in GR were considered.
We see that, when the star is described in terms of its mass, the regions of instability depend crucially on the equation of state.
Therefore, a precise measurement of only the mass of a neutron star can help us little in distinguishing GR from a STT, unless the underlying nuclear equation of state is well known. On the other hand, the dependence of the instability parameter space on the EOS diminishes a lot when the star is characterized by its compactness, as seen in the bottom panel of Fig.~\ref{fig:instability}. 
In particular, the minimum value of $M/R$ below which stars in each model have $T\leq 0$ everywhere (corresponding to the asymptotes in the bottom panel of Fig.~\ref{fig:instability}) is similar not only for the five EOS of Table \ref{table:EOS}, but for all the 15 $npe\mu$ models in Fig.~\ref{fig:Tmax} that allow for a positive value of $T_\textrm{max}$ [on average $(M/R)_\textrm{min}=0.263$]. 
Therefore, the observation of neutron stars with $M/R \gtrsim 0.27$ is already indicative of a highly stiff EOS, and can provide ways to distinguish GR from some STTs with $\beta_0>0$.

\section{Comments on scalarization and the end state of the instability}
\label{sec:scalarization}

Up to now, we have focused on the linear stability of equilibrium solutions in STTs describing a star in a constant scalar field environment. A natural question that follows is whether the linear instability found for certain configurations persists in a nonlinear analysis. For instance, is it possible that the star and the scalar field adjust themselves to a new stable configuration with lower total energy, dissipating the difference in energy by the emission, e.g., of gravitational waves? As a first step to answering that question, one can ask whether there exist other equilibrium solutions of Eqs.~(\ref{g_eq})-(\ref{phi_eq}), with the same baryon number as the unstable one, but a nontrivial profile for the scalar field. Indeed, such scalarized solutions do exist whenever the constant-scalar-field configuration is unstable \cite{Harada1998,Pani2011}. Yet, although much is known about scalarization when $\beta_0 < 0$, only a few works, most notably Ref.~\cite{Pani2011}, have addressed some features of the $\beta_0 > 0$ case. However, in that paper, the EOS was such that Eq.~(\ref{T}) was only satisfied inside stars which were already unstable in GR.
Therefore, in order to shed further light at this point, it is interesting to perform an analysis similar to that of Ref.~\cite{Pani2011} for (some of) the EOS used in Sec.~\ref{sec:paramspace}, for which the necessary condition for instability, namely $T>0$ inside the star, is achieved in configurations which are stable according to GR. 

For that purpose, we must first completely specify the STT, by choosing a particular form for $a(\phi)$ in Eq.~(\ref{general_action}). Note that higher order terms in Eq.~(\ref{a}) were irrelevant in the linear analysis, but must be accommodated in the general case. One simple choice would be $a(\phi) = \frac{\beta_0}{2} (\phi - \phi_0)^2$, but, rather, we will focus, as in Refs.~\cite{Salgado1998,Lima2010,Pani2011}, on the physically more interesting case of a (massless) nonminimally coupled scalar field $\bar{\phi}$, in which case we have $A(\bar{\phi}) = 1-8\pi\xi \bar{\phi}^2$, $B(\bar{\phi}) = 4\pi$, $V(\bar{\phi}) = 0$ and $a(\bar{\phi}) = 0$ in Eq.~(\ref{general_action}). The field equations then read
\begin{align} \label{g_eq2}
	\bar{G}_{\mu\nu} & = 8\pi  \left[ \bar{T}_{\mu\nu} + \bar{\nabla}_\mu \bar{\phi} \bar{\nabla}_\nu \bar{\phi} - (1/2) \bar{g}_{\mu\nu} \bar{\nabla}_\rho \bar{\phi} \bar{\nabla}^\rho \bar{\phi} \right. \nonumber \\
	& \left. -2\xi \bar{\nabla}_\mu (\bar{\phi} \bar{\nabla}_\nu \bar{\phi}) + 2\xi \bar{g}_{\mu\nu} \bar{\nabla}_\rho(\bar{\phi} \bar{\nabla}^\rho \bar{\phi}) \right ] (1-8\pi \xi \bar{\phi}^2)^{-1},
\end{align}
\begin{equation}\label{phi_eq2}
(- \bar{\nabla}_\mu \bar{\nabla}^\mu +\xi \bar{R}) \bar{\phi} = 0,
\end{equation}
where $\bar{T}^{\mu\nu}\equiv 2 (-\bar{g})^{-1/2} \delta S_m [\Psi_m ; \bar{g}_{\rho\sigma} ]/\delta \bar{g}_{\mu\nu}$ is the energy-momentum tensor of matter fields, and bars indicate we are no longer in the Einstein frame used so far.
This theory can be rewritten [through field redefinitions $\bar{\phi} \to \phi = f(\bar{\phi})$ and $\bar{g}_{\mu\nu} \to g_{\mu\nu} = e^{2b(\bar{\phi})} \bar{g}_{\mu\nu}$] in the Einstein frame \cite{Damour1996}, yielding, in particular, an $a(\phi)$ such that $\alpha_0 = 0$ and $\beta_0 = 2\xi$, with higher-order terms in Eq.~(\ref{a}) depending on the only free parameter $\xi$. Therefore, the previous analysis applies straightforwardly to this case if we set $\beta_0 = 2\xi$. 

\begin{figure}[b]
\includegraphics[width=8.5cm]{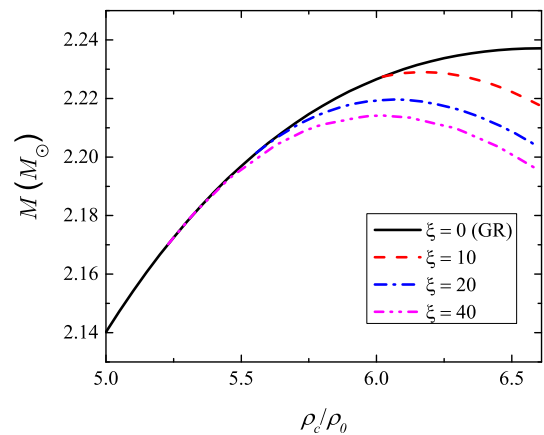}
\caption{Total gravitational mass as a function of the central rest-mass density (in units of $\rho_0 \equiv 2.7 \times 10^{14}$ g/cm$^3$), for  scalarized stellar equilibrium solutions with $\xi =$ 10, 20, and 40, as well as for neutron stars in GR. A piecewise polytropic approximation to the ENG model was employed.}
\label{fig:scalarization1}
\end{figure}

In the following, we adopt a similar procedure to Ref.~\cite{Pani2011} (see also Ref.~\cite{Salgado1998} for more details) in order to construct static spherically symmetric solutions of Eqs.~(\ref{g_eq2})-(\ref{phi_eq2}) consisting of a perfect fluid star described by a polytropic approximation to the ENG EOS (see Table I), and a nontrivial profile for the scalar field, i.e., in which $\bar{\phi}(r)|_{r=0} = \phi_c \neq 0$ and $\bar{\phi}(r)|_{r \to \infty} \to 0$ (note that $\bar{\phi} = 0$ corresponds to $\phi = \phi_0 = \textrm{const}$ in the associated Einstein frame). In Fig.~\ref{fig:scalarization1} we plot the total mass as a function of the central rest-mass density for sequences of equilibrium solutions with a nontrivial scalar field profile in theories with $\xi =$ 10, 20, and $40$. The general relativistic sequence, corresponding to the absence of the scalar field, is also represented. Note that the existence of a turning point criterion analogous to GR's in STTs \cite{Harada1998} indicates that scalarized equilibrium solutions in the low density branch before the turning point are thermodynamically stable. 
Most importantly, from Fig.~\ref{fig:scalarization1} we see that scalarized configurations have a lower maximum mass than in pure GR, and this maximum mass diminishes with increasing $\xi$. 

For the $\beta_0<0$ case, much is known about the evolution towards scalarization, both for static \cite{Novak1998,Alcubierre2010} and dynamical \cite{Barausse2013,Shibata2014,Palenzuela2014} initial states. However, the end state of the instability for theories with $\beta_0>0$ is still an open topic for numerical research, and it would be particularly interesting to investigate whether scalarized configurations could naturally arise from astrophysical processes such as neutron star inspirals.
Nonetheless, our results already indicate that, even when the scalar field is treated nonlinearly, the maximum mass (or compactness) of stable configurations in STTs with $\beta_0>0$ can be lower than in GR for some realistic EOS for neutron stars.

\section{Conclusions}
\label{sec:conclusions}

Scalar-tensor theories with $\alpha_0=0$ and $\beta_0>0$ are nowadays unconstrained by observations. A possibility of testing these theories arises if there are stable equilibrium solutions for matter fields in GR in which the trace $T$ of the energy-momentum tensor acquires positive values, since such configurations would be linearly unstable under scalar field perturbations in STTs with a large enough $\beta_0>0$ (cf. Sec.~\ref{sec:preliminars}). In fact, many equations of state for neutron stars which satisfy the current constraints allow for configurations which are stable in GR and obey this condition on $T$ (cf. Sec.~\ref{sec:trace}). In particular, this is true for several theoretical EOS, usually those which are stiffer in the stellar core and predict stars with compactness greater than $\sim 0.27$: this suggests that the observation of highly compact neutron stars could be used as a test of the underlying theory of gravity.

The parameter space of the instability was studied in Sec.~\ref{sec:paramspace}: there, we considered piecewise polytropic approximations to a set of realistic EOS and determined the ranges of stellar masses and compactnesses which would lead to an instability in STTs with various values of $\beta_0$. This parameter space is less EOS dependent when the stars are characterized by their compactnesses, as shown in Fig.~\ref{fig:instability}.
In Sec.~\ref{sec:scalarization}, we constructed static equilibrium solutions in STTs with $\beta_0>0$, describing neutron stars and a nontrivial scalar field profile, for a proper choice of EOS. In particular, we showed that these scalarized equilibrium configurations allow for a lower maximum mass than the GR solution (cf. Fig.~\ref{fig:scalarization1}). 

Some interesting questions are worthy topics of future research: what the end state is of the linear instability found for some stellar solutions in SSTs with $\beta_0>0$, and whether and how scalarized configurations form in these theories, specially in astrophysically motivated scenarios.
The answer to these questions, together with a better understanding of the equation of state of neutron stars, will be necessary in order to determine precisely whether and how neutron stars can be used to access this yet unconstrained region of the parameter space of STTs.

\acknowledgments
The author thanks George E. A. Matsas and Daniel A. T. Vanzella for the insightful discussions which led to this work, and Eric Poisson and Luis Lehner for important and helpful comments.
This work was supported by Conselho Nacional de Desenvolvimento Cient\'\i fico e Tecnol\'ogico (CNPq).


\begin{thebibliography}{}

\bibitem{Will2006} C. M. Will, Living Rev. Relativity \textbf{9}, 3 (2006).

\bibitem{Damour1992} T. Damour and G. Esposito-Far\`ese, Classical Quantum Gravity \textbf{9}, 2093 (1992).

\bibitem{Fujii2003} Y. Fujii and K. Maeda, \textit{The Scalar-Tensor Theory of Gravitation} (Cambridge University Press, Cambridge, 2003).

\bibitem{Sotiriou2008}  T. P. Sotiriou, S. Liberati, and V. Faraoni, Int. J. Mod. Phys. \textbf{D17}, 399 (2008).

\bibitem{Jordan1959} P. Jordan, Z. Phys. \textbf{157}, 112 (1959).

\bibitem{Brans1961} C. Brans and R. H. Dicke, Phys. Rev. \textbf{124}, 925 (1961).

\bibitem{Damour1993}  T. Damour and G. Esposito-Far\`ese, Phys. Rev. Lett. \textbf{70}, 2220 (1993).

\bibitem{Damour1996} T. Damour and G. Esposito-Far\`ese, Phys. Rev. D \textbf{54}, 1474 (1996).

\bibitem{Freire2012}  P. C. C. Freire, N. Wex, G. Esposito-Far\`ese, J. P. W. Verbiest, M. Bailes, B. A. Jacoby, M. Kramer, I. H. Stairs,
J. Antoniadis, and G. H. Janssen, Mon. Not. R. Astron. Soc. \textbf{423}, 3328 (2012).

\bibitem{Damour1996a} T. Damour and G. Esposito-Far\`ese, Phys. Rev. D \textbf{53}, 5541 (1996).

\bibitem{Bertotti2003}  B. Bertotti, L. Iess, and P. Tortora, Nature (London) \textbf{425}, 374 (2003).

\bibitem{Shibata2014} M. Shibata, K. Taniguchi, H. Okawa, and A. Buonanno, Phys. Rev. D \textbf{89}, 084005 (2014).

\bibitem{Doneva2013} D. D. Doneva, S. S. Yazadjiev, N. Stergioulas, and K. D. Kokkotas, Phys. Rev. D \textbf{88}, 084060 (2013).

\bibitem{Mendes2014b} R. F. P. Mendes, G. E. A. Matsas, and D. A. T. Vanzella, Phys. Rev. D \textbf{90}, 044053 (2014).

\bibitem{Harada1998}  T. Harada, Phys. Rev. D \textbf{57}, 4802 (1998).

\bibitem{Barausse2013}  E. Barausse, C. Palenzuela, M. Ponce, and L. Lehner, Phys. Rev. D \textbf{87}, 081506 (2013).

\bibitem{Palenzuela2014} C. Palenzuela, E. Barausse, M. Ponce, and L. Lehner, Phys. Rev. D \textbf{89}, 044024 (2014).

\bibitem{Harada1997} T. Harada, Prog. Theor. Phys. \textbf{98}, 359 (1997).

\bibitem{Haensel2007}  P. Haensel, A. Potekhin, and D. Yakovlev, \textit{Neutron Stars 1: Equation of State and Structure} (Springer, New York, 2007).

\bibitem{Damour1993b}  T. Damour and K. Nordtvedt, Phys. Rev. Lett. \textbf{70}, 2217 (1993).

\bibitem{Damour1993a}  T. Damour and K. Nordtvedt, Phys. Rev. D \textbf{48}, 3436 (1993).

\bibitem{Lima2010} 
W. C. C. Lima and D. A. T. Vanzella, 
Phys. Rev. Lett. \textbf{104}, 161102 (2010); 
W. C. C. Lima, G. E. A. Matsas, and D. A. T. Vanzella, 
Phys. Rev. Lett. \textbf{105}, 151102 (2010).

\bibitem{Mendes2014}  R. F. P. Mendes, G. E. A. Matsas, and D. A. T. Vanzella, Phys. Rev. D \textbf{89}, 047503 (2014).

\bibitem{Pani2011}
P. Pani, V. Cardoso, E. Berti, J. Read, and M. Salgado, 
Phys. Rev. D \textbf{83}, 081501 (2011).

\bibitem{Novak1998}
J. Novak, Phys. Rev. D \textbf{58}, 064019 (1998).

\bibitem{Zeldovich1962}  Y. B. Zeldovich, J. Exp. Theor. Phys. \textbf{14}, 1143 (1962).

\bibitem{Koranda1997}  S. Koranda, N. Stergioulas, and J. L. Friedman, Astrophys. J. \textbf{488}, 799 (1997).

\bibitem{Sorkin1981} R. D. Sorkin, Astrophys. J. \textbf{249}, 254 (1981).

\bibitem{Lindblom1992}  L. Lindblom, Astrophys. J. \textbf{398}, 569 (1992).

\bibitem{Steiner2010}  A. W. Steiner, J. M. Lattimer, and E. F. Brown, Astrophys. J. \textbf{722}, 33 (2010).

\bibitem{Read2009}  J. S. Read, B. D. Lackey, B. J. Owen, and J. L. Friedman, Phys. Rev. D \textbf{79}, 124032 (2009).

\bibitem{Douchin2001}  F. Douchin and P. Haensel, Astron. Astrophys. \textbf{380}, 151 (2001).

\bibitem{Demorest2010}  P. B. Demorest, T. Pennucci, S. M. Ransom, M. S. E.
Roberts, and J. W. T. Hessels, Nature (London) \textbf{467}, 1081 (2010).

\bibitem{Antoniadis2013} J. Antoniadis, P. C. C. Freire, N. Wex, T. M. Tauris, R. S. Lynch, M. H. van Kerkwijk, M. Kramer, C. Bassa, V. S.
Dhillon, T. Driebe, J. W. T. Hessels, V. M. Kaspi, V. I. Kondratiev, N. Langer, T. R. Marsh, M. A. McLaughlin, T. T. Pennucci, S. M. Ransom, I. H. Stairs, J. van Leeuwen, J. P. W. Verbiest, and D. G. Whelan, Science
\textbf{340}, 1233232 (2013).

\bibitem{Salgado1998}  M. Salgado, D. Sudarsky, and U. Nucamendi, Phys. Rev. D \textbf{58}, 124003 (1998).

\bibitem{Alcubierre2010} M. Alcubierre, J. C. Degollado, D. N\'{u}\~nez, M. Ruiz, and M. Salgado, Phys. Rev. D \textbf{81}, 124018 (2010).




\end{thebibliography}

\end{document}